\documentclass[%
 reprint,
superscriptaddress,
 amsmath,amssymb,
 aps,
]{revtex4-2}
\usepackage{graphicx}
\usepackage{dcolumn}
\usepackage{color}
\usepackage{comment}
\usepackage{multirow}
\usepackage{hyperref}
\usepackage{float}
\makeatletter
\let\newfloat\newfloat@ltx
\makeatother
\usepackage{algorithm}
\usepackage{algpseudocode}
\usepackage{bm}
\usepackage{mathrsfs}
\usepackage{float}
\usepackage{amsmath}
\usepackage{warpcol}
\usepackage{longtable}
\graphicspath{{Images/}}
\usepackage{xcolor}
\definecolor{red1}{rgb}{0.6,0,0}

\usepackage{longtable,booktabs}

\begin{document}

\title{A general optimization framework for mapping local transition-state networks}

\newcommand{\KTH}{Department of Applied Physics, School of Engineering Sciences, KTH Royal Institute of Technology, 
AlbaNova University Center, SE-10691 Stockholm, Sweden}

\newcommand{\SeRC}{Swedish e-Science Research Center, KTH Royal Institute of Technology, SE-10044 Stockholm, Sweden}

\newcommand{\WISEKTH}{Wallenberg Initiative Materials Science for Sustainability (WISE), KTH Royal Institute of Technology, SE-10044 Stockholm, Sweden}

\author{Qichen Xu*}
    \affiliation{\KTH}
    \affiliation{\SeRC}
    \thanks{qichenx@kth.se}

\author{Anna Delin}
    \affiliation{\KTH}
    \affiliation{\SeRC}
        \affiliation{\WISEKTH}

\date{\today}

\begin{abstract}
Understanding how complex systems transition between states requires mapping the energy landscape that governs these changes.
Local transition-state networks reveal the barrier architecture that explains observed behaviour and enables mechanism-based prediction across computational chemistry, biology, and physics, yet current practice either prescribes endpoints or randomly samples only a few saddles around an initial guess. We present a general optimization framework that systematically expands local coverage by coupling a multi-objective explorer with a bilayer minimum-mode kernel.  The inner layer uses Hessian–vector products to recover the lowest-curvature subspace (smallest $k$ eigenpairs), the outer layer optimizes on a reflected force to reach index-1 saddles, then a two-sided descent certifies connectivity. The GPU-based pipeline is portable across autodiff backends and eigensolvers and, on large atomistic-spin tests, matches explicit-Hessian accuracy while cutting peak memory and wall time by orders of magnitude. Applied to a DFT-parameterized Néel-type skyrmionic model, it recovers known routes and reveals previously unreported mechanisms, including meron–antimeron–mediated Néel-type skyrmionic duplication, annihilation, and chiral-droplet formation, enabling up to 32 pathways between biskyrmion ($Q=2$) and biantiskyrmion ($Q=-2$). The same core transfers to Cartesian atoms, automatically mapping canonical rearrangements of a Ni(111) heptamer, underscoring the framework’s generality.

\end{abstract}

\maketitle

\section{Introduction}
\begin{figure*}
    \centering
    \includegraphics[width=18cm]{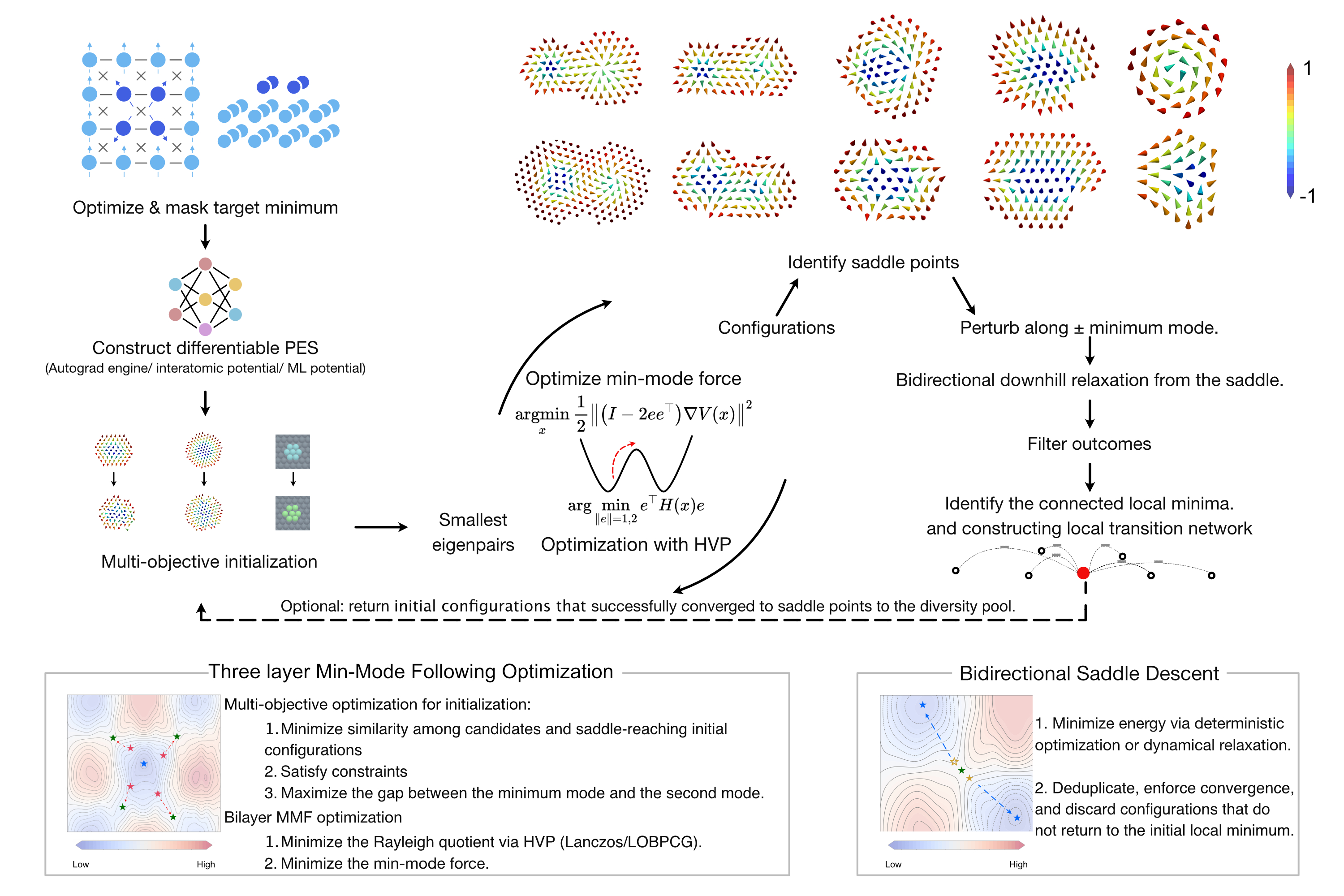}
    \caption{\textbf{Overview of the MOTO framework and optimization objectives.}
    Top: three-layer workflow: First layer, a multi-objective explorer proposes diverse, feasible initial guess. Second layer, a bilater minimum-mode kernel (direction optimization + ascent) locates index-1 saddles. Third layer, deterministic ±min-mode saddle descent certifies the transition states. Representative isolated transition states found around a Néel skyrmion and an antiskyrmion in DFT-parameterized Pd/Fe/Ir(111) are shown. Bottom: objectives and strategy used during optimization, which guide automated discovery of the local saddle–minima network.}
        \label{fig:fig1_workflow}
    \end{figure*}

The behavior of matter at the atomic and molecular scale is governed by intricate interactions that shape its structure, dynamics, and reactivity. These intricate interactions are encoded into the energy landscape, or potential energy surface.
Understanding the energy landscape, especially how the local minima are connected, of complex physicochemical systems remains a central challenge in condensed-matter physics and chemical science and new computational methods in this area have the potential to transform catalyst design for green chemistry, drug discovery and spintronics engineering.\cite{Wales2006JPCB,Wales2018ARPC,BinderYoung1986RMP,Onuchic1997ARPC,Bussi2020NatRevPhys,Shires2021PRX}
Local minima, in combination with the index-1 saddle points (transition states) that connect them, govern dynamics ranging from atomic diffusion and chemical reactions to thermally activated transformations of topological spin textures.\cite{Wales2006JPCB,Henkelman1999JCP_dimer,Henkelman2000JCP_CINEB,Bessarab2015CPC}

A common approach represents the global surface as a kinetic transition-state network and visualizes it with disconnectivity graphs or manifold-learning embeddings.\cite{Wales2006JPCB,Becker1997JCP,Ceriotti2011PNAS,Shires2021PRX} While powerful for revealing the distribution of stable structures, these tools often provide limited mechanistic insight and do not directly enable rate modeling. In practical simulations, two families dominate local transition-state searches for nearby index-1 saddles. Path-based methods, NEB/CI-NEB and, for magnetic systems, GNEB, optimize a discrete chain of images between prescribed endpoints to obtain a minimum-energy path, the highest-energy image (the band configuration of maximal energy, explicitly pushed to the first-order saddle in CI-NEB/GNEB) converges to the transition state.\cite{Henkelman2000JCP,Bessarab2015CPC} These methods are robust when initial and final states are known, but coverage is constrained by the need for informed endpoints. Single-ended minimum-mode-following (MMF) methods, such as the dimer method and subsequent variants, start near a minimum, seek the lowest-curvature direction, invert the force along that direction, and ascend to an index-1 saddle using only first derivatives.\cite{Henkelman1999JCP,Heyden2005JCP} Despite many refinements, building local networks reliably at scale remains difficult, systematic coverage of nearby saddles with streamlined workflows for large systems remains challenging\cite{schrautzer2025identification}.

Building on these opportunities, we introduce a multi-objective three-layer optimization (MOTO) framework for systematic local-landscape mapping: (i) a multi-objective explorer that proposes diverse, feasible initial guesses; (ii) a bilayer minimum-mode kernel: inner HVP-based direction optimization with outer minimum-mode-force ascent, to locate index-1 saddles, and (iii) a deterministic two-sided minimum-mode descent (±MM-Descent) to certify local connectivity. We first focus on atomistic spin systems for benchmarking, then assess portability on a Cartesian atomic benchmark. Within spin models, our implementation scales to $10^6$ spins while computing the two lowest curvature modes with less than 1~GB incremental memory and 0.5~s wall time. On a $70\times70$ spin system (the largest tractable for an explicit-Hessian solver on a single GPU), the HVP kernel reduces incremental peak memory to $\sim\!0.01\%$ of the explicit-Hessian approach and delivers a $\times10$ speedup on identical hardware. We demonstrate generality by mapping transition-state networks in an artificial Bloch-type atomistic-spin model, a DFT-parameterized Pd/Fe/Ir(111) model,\cite{miranda2022band,xu2023metaheuristic,xu2025design,xu2022genetic} and, as a Cartesian case, a Ni(111) heptamer.\cite{Heinze2011NatPhys,Romming2013Science} The workflow recovers established collapse and escape mechanisms in the artificial model and reveals previously unreported saddle families, including meron–antimeron–mediated duplication of N\'eel-type skyrmions, formation of chiral droplets, and a boundary-defect–induced change in topological charge. These transition states form a local network that enables up to 32 pathways between a biskyrmion ($Q=2$) and a bi-antiskyrmion ($Q=-2$), suggesting actionable routes for skyrmion creation and transportation in spintronics devices.
\section{Results}

\subsection{MOTO workflow}

\begin{figure*}
    \centering
    \includegraphics[width=18cm]{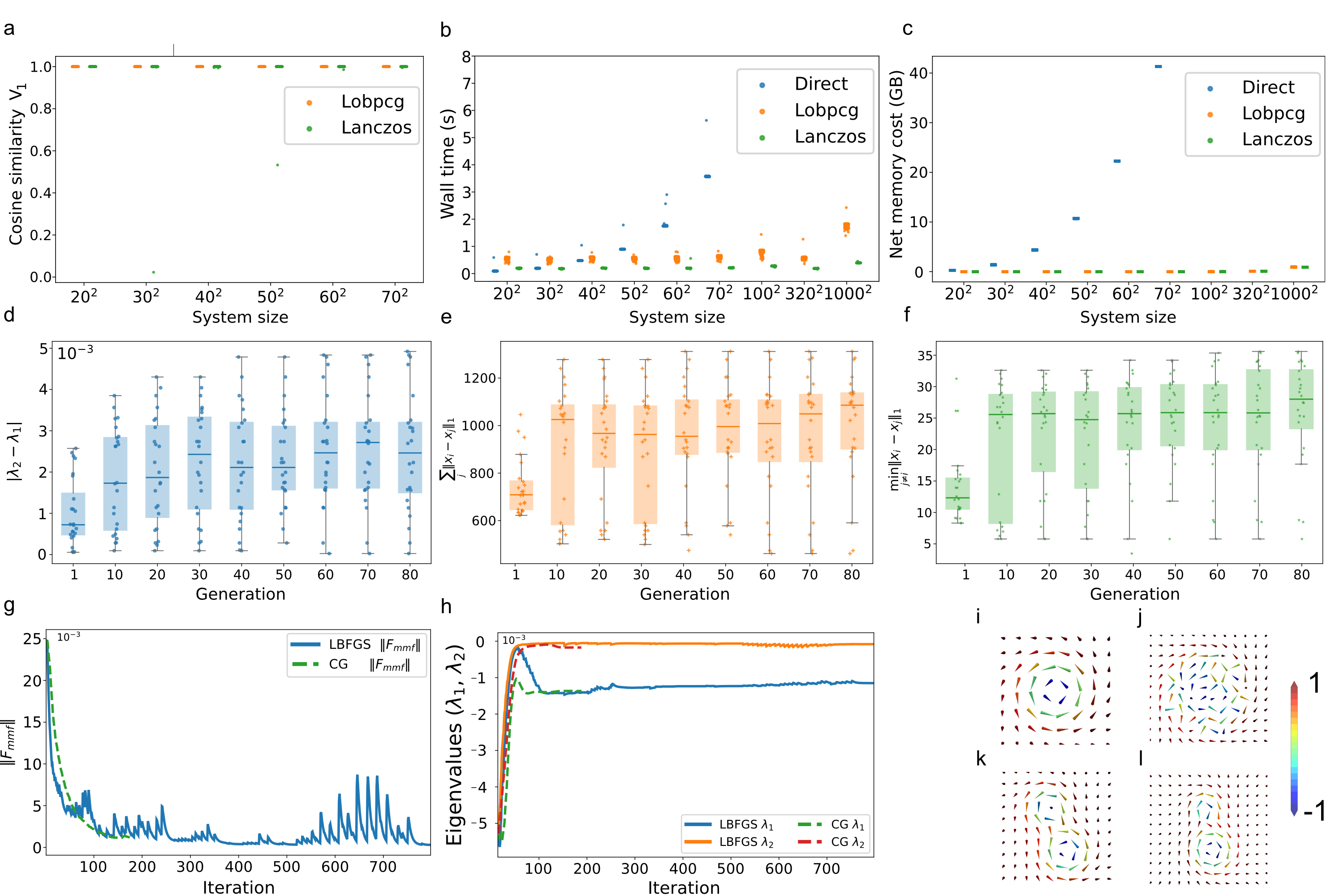}
    \caption{\textbf{Performance of the MOTO framework on artificial atomistic-spin system.} a–c Inner kernel (bilatet MMF) benchmarks. a, Cosine similarity of the minimum-mode eigenvector obtained with HVP–LOBPCG and HVP–Lanczos, measured against the reference eigenvector from the explicit Hessian, for different system sizes. b, Wall-time per minimum-mode update versus system size for the explicit-Hessian method, HVP–LOBPCG, and HVP–Lanczos. c, Peak memory versus system size for the same three solvers. Each point cluster in a–c aggregates 100 runs on random spin configurations. d–f Multi-objective explorer (population size 24). Objective values over generations using eigenvalues from Lanczos: d, spectral gap between the lowest and second-lowest tangent-space eigenvalues, box-and-whisker overlays show median, interquartile range, and 1.5× inter-quartile range whiskers. e and f, Evolution of the diversity objectives: total pairwise L1 distance (L1) and Maximin (minimum-distance) diversity (L1). g and h, Outer kernel (bilayer MMF) benchmarks. Convergence of the minimum-mode force (g) and the two lowest eigenvalues (h) under L-BFGS and conjugate-gradient (CG), showing approach to an index-1 saddle. For readability, traces in (g) are smoothed with an exponential moving average (EMA)\cite{Holt2004IJF}, the same smoothing convention widely used for training-loss curves in ML dashboards (e.g., TensorBoard). i and j, Initial spin configuration and the masked perturbation used in g–h. k and l, Transition states obtained from the bilayer MMF optmization with L-BFGS (k) and CG (l), respectively. All calculations are performed on Nvidia GH200 platfrom}
    \label{fig:fig2_performance}
    \end{figure*}

\begin{figure*}
    \centering
    \includegraphics[width=18cm]{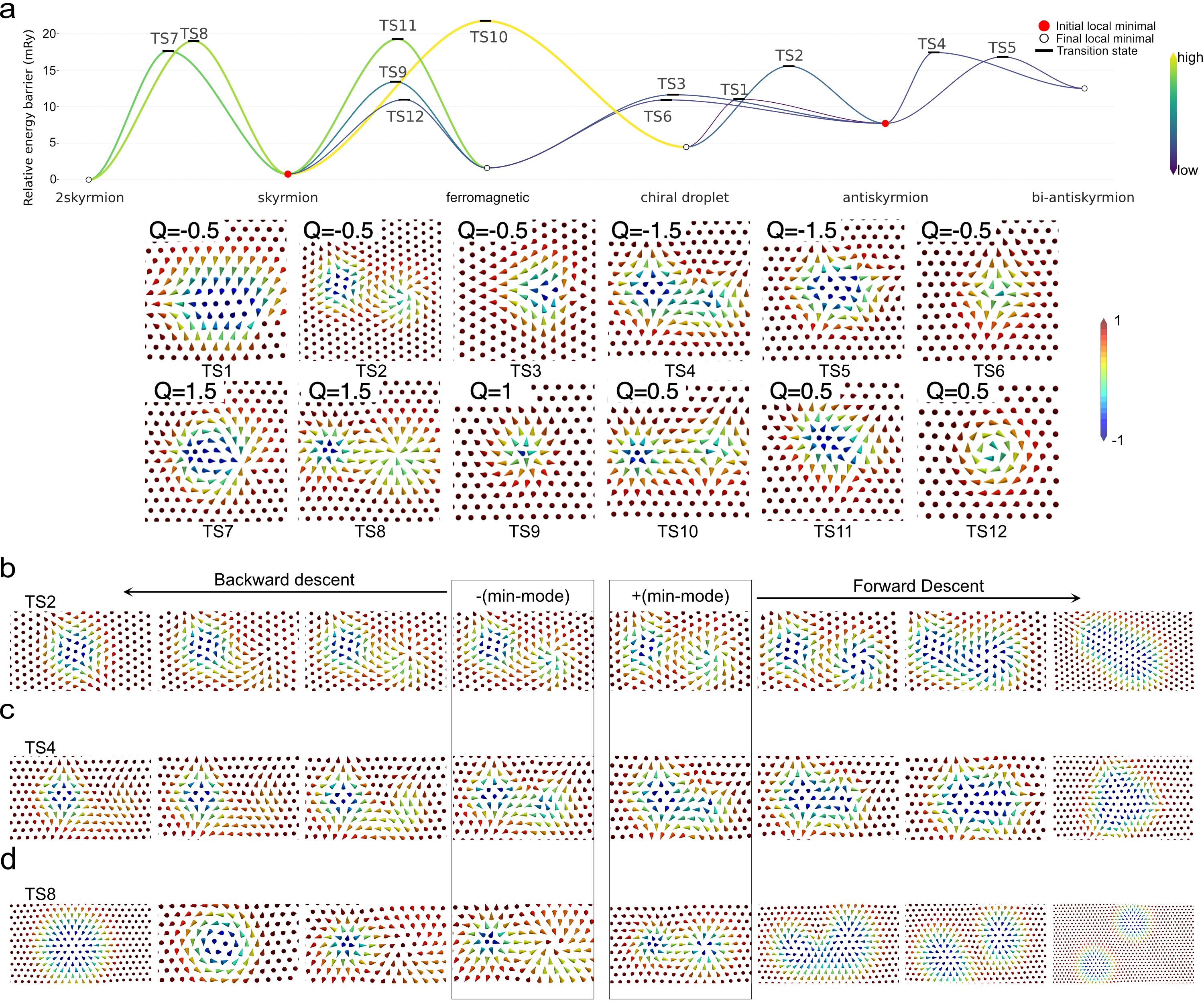}
    \caption{\textbf{Local transition networks of a Néel skyrmion and an antiskyrmion in a DFT-parameterized Pd/Fe/Ir(111) system system size 100X100 spins with external mangetic filed B=2.7T.} 
    a, Transition-state network starting from a Néel skyrmion (left) and an antiskyrmion (right). Bottom: thumbnails of the transition states referenced in the network. Final target local minima highlighted at the top of panel a, a chiral droplet, a bi-antiskyrmion, and two skyrmions, correspond to the right-hand endpoints of panels b, c, and d, respectively.
    b, Bidirectional saddle descent from TS2 (antiskyrmion–meron transition state): ±min-mode downhill trajectory snapshots.
    c, Bidirectional saddle descent from TS4 (antiskyrmion–antimeron transition state).
    d, Bidirectional saddle descent from TS9 (skyrmion–meron transition state). 
    All bidirectional saddle descent from TS1 to Ts12 are shown in support movie 1 to 12.
    }
        \label{fig:fig3_sknetworks}
    \end{figure*}

\begin{figure*}
    \centering
    \includegraphics[width=18cm]{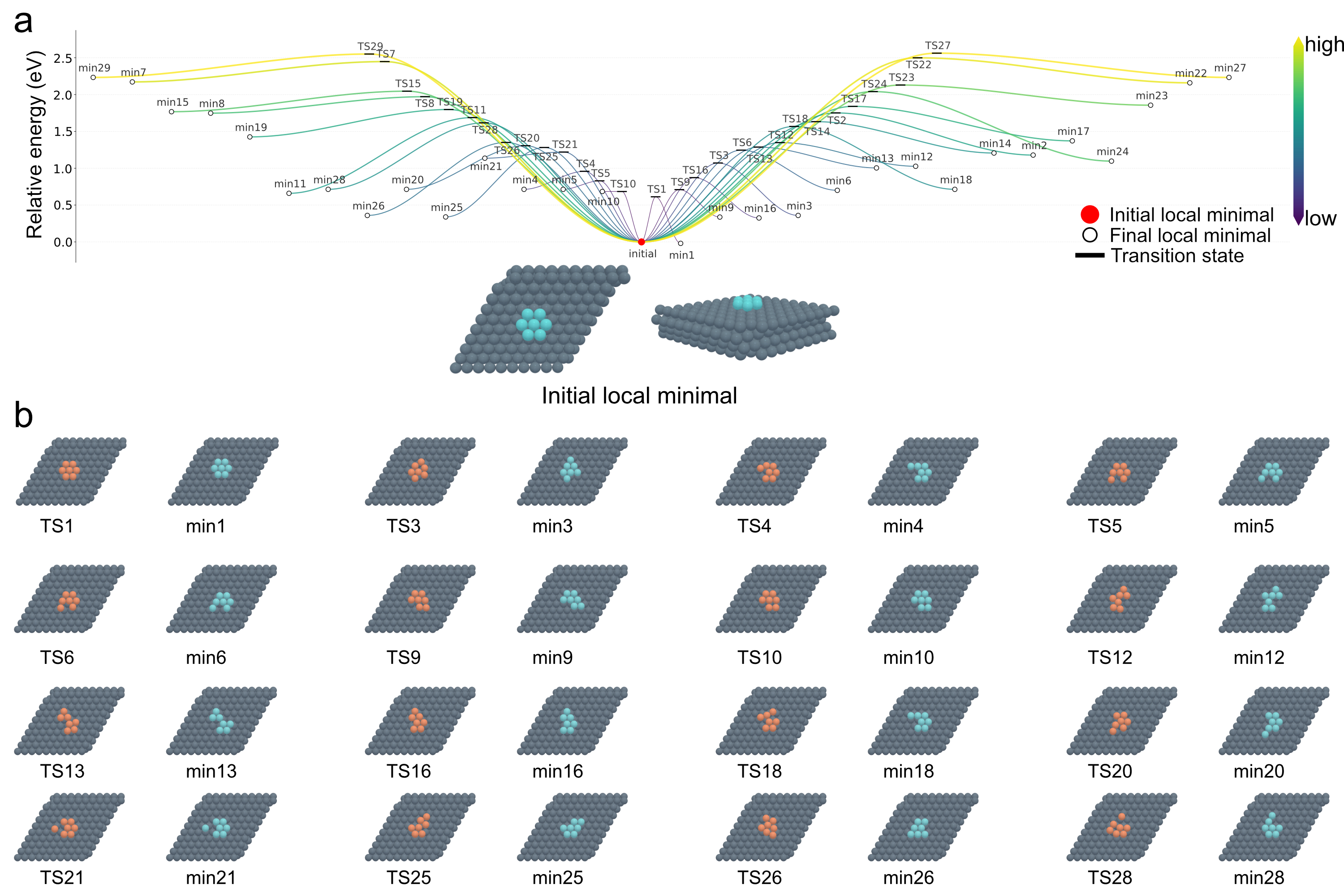}
    \caption{\textbf{Local transition-state landscape of a Ni heptamer on Ni(111).} a, Transition-state network mapped from a seven-atom Ni island placed on a four-layer fcc(111) slab (100 Ni atoms per layer). Bottom of a: top and side views of the initial heptamer configuration.
    b, Top views of low-barrier saddles ($\Delta E < 1.6\,\text{eV}$) and their corresponding final states.}
        \label{fig:fig4}
    \end{figure*}

\paragraph*{Multi-objective initial-configuration explorer.}
To construct the local transition-state network around a given minimum, we optimize a set of
masked perturbations (top-left panel of Fig.~\ref{fig:fig1_workflow}) so as to expose as many
distinct nearby index-1 saddles as possible. Let $\Omega\subset\{1,\dots,N\}$ denote the index set of
active degrees of freedom (the mask), and let $P_\Omega\in\mathbb{R}^{N\times|\Omega|}$ be the
selector. An initial configuration is generated as
\begin{equation}
x \;=\; x_0 \;+\; P_\Omega\,\delta, \qquad \delta\in\mathbb{R}^{|\Omega|},
\end{equation}
optionally followed by projection to the admissible manifold (e.g., spin–tangent retraction in
atomistic spin systems). We search the masked subspace with an evolutionary multi-objective scheme\cite{Deb2001Book}.
Specifically, NSGA-II evolves the perturbation vector $\delta$ via Pareto selection, yielding diverse and feasible proposals while suppressing trivial bulk motions and focusing exploration where changes are allowed\cite{Deb2002NSGAII,Blank2020pymoo}.

We use three objectives (bottom panel of Fig.~\ref{fig:fig1_workflow}):

(i) Diversity, minimize similarity among candidates and to an archive of previously
saddle-reaching initial configurations to maximize coverage in configuration space.

(ii) Feasibility, satisfy physical and geometric constraints encoded by normalized
violations (e.g., initial topological charge in atomistic spin systems, minimum interatomic distance in crystalline systems).

(iii) Minimum-mode separability, maximize the spectral gap $\Delta\lambda=\lambda_2-\lambda_1$ between the two smallest curvature modes (estimated via
HVP based Lanczos/LOBPCG)\cite{Lanczos1950,Knyazev2001LOBPCG}, which stabilizes minimum-mode estimation
and the subsequent ascent. 

NSGA-II returns a Pareto front $\mathcal{P}$ of initial configurations,
from each $x\in\mathcal{P}$ we estimate the minimum mode $\mathbf{u}^*(x)$ and launch Bilayer MMF.

\paragraph*{Bilayer MMF.}
Bilayer MMF abstracted minimum–mode following as two coupled optimizations.
Inner layer solves a spectral subproblem to identify the lowest–curvature subspace.
Using HVP–driven Krylov eigensolvers (Lanczos or LOBPCG) in operator mode, we minimize the Rayleigh
quotient to obtain the smallest $k$ eigenpairs of the masked tangent Hessian, yielding
$V_{\min}(x)=[\mathbf v_1,\dots,\mathbf v_k]$ and $\{\lambda_1,\dots,\lambda_k\}$.
For well separated minima we set $k=1$ and write $\mathbf v_{\min}(x)=\mathbf v_1$.

Outer layer then optimizes on the minimum–mode force field to reach an index-1 saddle.
Given $\mathbf v_{\min}(x)$, the update uses
\begin{equation}
\mathbf F_{\mathrm{mmf}}(x)
=
\mathbf F(x)
-2\bigl[\mathbf F(x)^{\!\top}\mathbf v_{\min}(x)\bigr]\mathbf v_{\min}(x),
\end{equation}
we climb with a GPU-based optimizer (LBFGS or CG) with line search.
When $k>1$ we use the subspace variant by replacing $\mathbf v_{\min}(x)\mathbf v_{\min}(x)^{\!\top}$ with
$P_{\min}(x)=V_{\min}(x)V_{\min}(x)^{\!\top}$.
Both layers are modular, any eigensolver that returns the $k$ smallest modes and any smooth optimizer
on $\mathbf F_{\mathrm{mmf}}$ can be swapped in without changing the framework logic. More details about the computational pipelines are described in the Methods section and Support Note 1.  

\paragraph*{Deterministic bidirectional saddle decent} Once the saddle point is located, we perturb
along $\pm\,\mathbf{v}_{\min}(x)$ and relax bidirectionally to certify connectivity, deduplicate
outcomes, and add the resulting minima and their associated initial configurations to the archive.
This explore $\rightarrow$ ascend $\rightarrow$ certify $\rightarrow$ archive loop expands the
local transition-state network outward from the reference minimum with high coverage and low redundancy.

\subsection{Computational performance analysis}

As shown in Fig.~\ref{fig:fig2_performance}, we benchmark the three components of MOTO on atomistic–spin
testbeds, exchange paramters can be found in UppASD repository, focusing on (i) accuracy and cost of the inner kernel (minimum-mode estimation),
(ii) the behavior of the multi-objective explorer over generations, and (iii) convergence of the
outer MMF optimization kernel.

\paragraph*{Inner kernel: HVP based optimization vs. explicit Hessian.}
Figure~\ref{fig:fig2_performance}a–c compares HVP based Lanczos/LOBPCG with an explicit-Hessian
baseline across system sizes. To control variables, the maximum iteration of the Lanczos/LOBPCG
optimizers is capped at 200. Both Krylov solvers recover the minimum-mode direction with
near-unity cosine similarity relative to the reference eigenvector from the explicit Hessian
(Fig.~\ref{fig:fig2_performance}a), confirming that HVP based estimation in the masked tangent
space is accurate (two outliers are observed for Lanczos). In wall time, the HVP based solvers scale
gently with size, whereas the explicit-Hessian method grows much more steeply
(Fig.~\ref{fig:fig2_performance}b), in particular, Lanczos performs better at larger sizes, so we
use it for the remaining experiments. In peak net memory, the contrast is more pronounced:
explicit-Hessian construction has a rapidly increasing footprint, while HVP based
Lanczos/LOBPCG exhibit nearly size-invariant incremental peak memory
(Fig.~\ref{fig:fig2_performance}c). Each point cluster aggregates 100 random initial configurations
under identical hardware and backend, reported memory is the incremental peak relative to the
autograd baseline.

\paragraph*{Multi-objective explorer over generations.}
With a population size of 24, Fig.~\ref{fig:fig2_performance}d–f tracks the three objectives using
eigenvalues from Lanczos. The spectral gap $\Delta\lambda=\lambda_2-\lambda_1$ steadily increases
with generation (Fig.~\ref{fig:fig2_performance}d), improving minimum-mode separability and
stabilizing downstream ascent. Diversity also improves: the total pairwise $\ell_1$ distance
(Fig.~\ref{fig:fig2_performance}e) and the maximin (minimum-distance) diversity in $\ell_1$
(Fig.~\ref{fig:fig2_performance}f) both grow and then plateau, indicating wide coverage with reduced
redundancy. Box-and-whisker overlays show the median, interquartile range, and 1.5$\times$ IQR
whiskers.

\paragraph*{Outer kernel (Bilayer MMF) convergence.}
Figure~\ref{fig:fig2_performance}g–h assesses the outer ascent using L-BFGS and conjugate gradient
(CG). The norm of the minimum-mode force $\|\mathbf{F}_{\mathrm{mmf}}\|$ decreases by orders of
magnitude (Fig.~\ref{fig:fig2_performance}g), while the two smallest curvature eigenvalues approach
the index-1 signature with $\lambda_1<0<\lambda_2$ (Fig.~\ref{fig:fig2_performance}h).
Figures~\ref{fig:fig2_performance}i–j show a representative initial configuration and its masked
perturbation, Figs.~\ref{fig:fig2_performance}k–l display the resulting transition states from
Bilayer MMF with L-BFGS and CG, respectively. Both optimizers reach an index-1 saddle within a
reasonable number of iterations, small differences in trajectories reflect line-search and
preconditioning effects (CG tends to stop earlier, while L-BFGS attains a slightly lower
$\|\mathbf{F}_{\mathrm{mmf}}\|$). The illustrated case corresponds to an elongation-driven duplication
path in the toy Bloch-type model.

In summary, HVP based optimization methods match explicit-Hessian accuracy while dramatically reducing memory and runtime within autograd engine,  the multi-objective explorer reliably increases minimum-mode separability and diversity, expanding the effective exploration region and the Bilayer MMF outer ascent converges robustly with distinct optimizers, certifying index-1 saddles and closing the loop for automated construction of the local transition-state network.

\subsection{Skyrmion/antiskyrmion seeded local transition-state networks in Pd/Fe/Ir(111)} 
We applied MOTO to map the local transition-state network around a N\'eel skyrmion (\(Q=+1\)) and an antiskyrmion (\(Q=-1\)) in a DFT-parameterized Pd/Fe/Ir(111) model at \(B=2.7\,\mathrm{T}\). Simulation box is 100X100 spins with a periodic boundary condition.
The resulting network and sample transition states are shown in Fig.~\ref{fig:fig3_sknetworks}a.
In this neighborhood we find twelve different index-1 saddles (TS1 to TS12).
Many of them show boundary-defect and affiliated meron/antimeron, carrying half-integer skyrmion charge (e.g., \(Q=\pm 0.5,\,1.5\)), consistent with meron/antimeron segments pinned at the texture edge or at a boundary defect.
The network connects the reference minima to nearby basins, a chiral droplet, a bi-antiskyrmion, and a two-skyrmion state, through up to 32 distinct transition channels (Fig.~\ref{fig:fig3_sknetworks}a).

Bidirectional saddle descents (Fig.~\ref{fig:fig3_sknetworks}b–d) certify the connectivity and clarify the underlying mechanisms. From TS2, identified as an antiskyrmion–meron saddle, the negative minimum-mode direction returns to the antiskyrmion basin while the positive direction generates a chiral droplet (Fig.~\ref{fig:fig3_sknetworks}b). From TS4, an
antiskyrmion–antimeron saddle, the two directions connect the antiskyrmion and a bi-antiskyrmion minimum (Fig.~\ref{fig:fig3_sknetworks}c). From TS8, a skyrmion–meron saddle, the forward descent produces two skyrmions via an elongation-driven duplication, while the backward descent returns to the single skyrmion
(Fig.~\ref{fig:fig3_sknetworks}d). These paths exemplify a consistent picture in which perimeter-anchored meron/antimeron and defects mediate charge transfer between basins.

Taken together, the mapped network highlights two salient features. First, many transition states induced by meron or antimeron, supplying local transition mechanisms in the neighbourhood of both Néel-type skyrmions and antiskyrmions. Second, the downhill outcomes follow simple $\Delta Q$
 selection rules consistent with boundary-defect accounting: meron-mediated branches change the global skyrmion number by $\pm 1$, whereas paired meron/antimeron segments move between neighbouring basins without spurious intermediate trapping. Obtained automatically from a single MOTO run, these results provide broad coverage of nearby saddles and a clear mechanistic picture of how duplication, annihilation, and chiral-droplet formation emerge in this material system.

\subsection{Local transition-state networks in the Ni(111) heptamer}

As noted above, the approach is flexible: it is not limited to atomistic spin models and applies to any
differentiable potential energy surface(PES). Here we apply MOTO to a classical surface-diffusion benchmark—a seven-atom Ni
island on Ni(111). The model uses a four-layer fcc(111) slab (100 Ni atoms per layer), island atoms
are treated as active degrees of freedom in Cartesian coordinates, while the bottom layers are fixed.
The workflow is unchanged across domains: the multi-objective explorer operates on masked Cartesian
perturbations, the inner layer estimates the minimum mode via HVP-based Lanczos/LOBPCG, and the outer
layer climbs using the minimum-mode force. The PES is given by a standardized EAM potential\cite{foiles1986embedded}.

Figure~\ref{fig:fig4}a shows the local transition-state network constructed around the compact
heptamer minimum (insets: top and side views). The network contains a dense set of nearby index-1
saddles and their connected minima on both sides of the reference basin. Within the barrier window
highlighted in Fig.~\ref{fig:fig4}b ($\Delta E<1.6$ eV), the workflow recovers the canonical
rearrangement motifs for this benchmark—edge and corner events as well as multi-atom concerted
moves—and certifies their connectivity by bidirectional descent (each saddle thumbnail in
Fig.~\ref{fig:fig4}b is paired with its relaxed final state). In partular, some of them are vaildated by machine learning potential MACE and Fairchem but since machine learning potentials still fast-developing research frontiers and certainly not perfect at present, not all saddle points can be found by them.

The key point of this benchmark is generality rather than new physics: without any change to the
algorithmic core, MOTO transfers from spin manifolds to atomic Cartesian systems, automatically maps
a rich local transition-state network from a single minimum, and produces inputs suitable for
downstream kinetic modeling. It is worth noting that when HVPs are unavailable or impractical (e.g., black-box DFT evaluations), the inner layer of the HVPs-based optimization can be replaced by force-based bilayer MMF optimization kernel such as the dimer method, which often yield better performance.

\subsection{Discussion and outlook}     

We presented MOTO, a complete optimization framework for mapping local transition-state networks around a given minimum. The workflow has three essential components: (i) a multi-objective initial-configuration explorer that balances diversity, feasibility, and minimum-mode separability, (ii) an HVP-based Bilayer MMF kernel that estimates the minimum-curvature subspace with Lanczos/LOBPCG and climbs using the minimum-mode force, and (iii) a deterministic two-sided descent to certify connectivity. Unlike single-path or single-seed strategies, MOTO explicitly optimizes a set of starting configurations and couples them to a scalable minimum-mode kernel, delivering systematic local coverage with modest memory and wall-time.

Evidence for generality comes from two domains. In a DFT-parameterized Pd/Fe/Ir(111) spin model, MOTO reconstructed a rich local network around Néel skyrmion and antiskyrmion minima and uncovered multiple meron/antimeron-mediated routes, including duplication, vanishing and droplet formation. In a classical surface-diffusion benchmark—a Ni heptamer on Ni(111)—the same algorithmic core, with only the projector switched to Cartesian updates, recovered the canonical set of low-barrier rearrangements and their connected minima. No hand-crafted moves or method retuning were required.

Limitations remain. The mask and the multi-objective targets must match the physics of the system and the region where changes are allowed. Performance is sensitive to kernel hyperparameters such as Krylov depth, reorthogonalization, and line-search thresholds. Poor settings can reduce diversity or stall the outer ascent. Near-degenerate curvature can also mislead a single direction. Robustness can be improved by following a low-dimensional minimum-eigenspace and by adding preconditioning to the Krylov solvers.

Looking forward, two directions are especially promising for further works: (i) automation—principled hyperparameter tuning and stopping rules tied to coverage metrics, and (ii) interoperation—tight coupling with two-end path solvers for kinetic-pathway identification. By combining multi-objective exploration with HVP-based minimum-mode following in a single loop, MOTO offers a practical route to high-coverage, mechanism-aware local landscape maps across differentiable PESs,from spin manifolds to atomic Cartesian systems—while remaining compatible with established transition-state search paradigms.

\section{Methods}
\subsection{Spin Hamiltonian and manifold gradient}


We consider an atomistic spin system with $N$ unit-length spins, each represented by a three-dimensional vector $\mathbf m_i\in\mathbb R^3$ with $\|\mathbf m_i\|_2=1$. 
Stacking all spins yields the configuration vector
\begin{equation}
    \mathbf m
    \;=\;
    \begin{bmatrix}
        \mathbf m_1^\top & \mathbf m_2^\top & \cdots & \mathbf m_N^\top
    \end{bmatrix}^{\!\top}
    \in \mathbb R^{3N},
\end{equation}
where ${}^\top$ denotes transpose and $\mathbf m_i$ is the normalized spin at site $i$.

The total energy $E(\mathbf m)$ includes exchange, Dzyaloshinskii–Moriya interaction (DMI), Zeeman coupling, and uniaxial anisotropy.
We write the Hamiltonian as $E(\mathbf m)$ (rather than the customary $\mathcal{H}$ or $H$) to avoid clashes with the Hessian notation $H(x)=\nabla^2 E(x)$ used in the optimization algorithms.

\begin{equation}\label{eq:param-hamiltonian}
\begin{aligned}
E(\mathbf m)
&=
-\sum_{i \neq j} J_{i j} \mathbf{m}_{i} \cdot \mathbf{m}_{j}-
\sum_{i \neq j} \mathbf{D}_{i j}\cdot\left(\mathbf{m}_{i} \times \mathbf{m}_{j}\right) \\&- \sum_{i} \mu_{i}\mathbf{B}^{\mathrm{ext}} \cdot \mathbf{m}_{i}  - \sum_{i}K^{\mathrm{U}}_i\left(\mathbf{m}_{i} \cdot \mathbf{e}_{z}\right)^{2},
\end{aligned}
\end{equation}
Here $J_{ij}\in\mathbb{R}$ is the Heisenberg exchange, $\mathbf D_{ij}\in\mathbb{R}^3$ the DMI vector,
$\mathbf e_z$ the easy-axis direction, and $K^{\mathrm{U}}_i$ the uniaxial anisotropy strength for site $i$.

\subsection{Gradient, HVP, and Hessian via the autograd engine}

The total spin energy is a differentiable scalar $E:\mathbb{R}^{3N}\to\mathbb{R}$ on
$\mathbf m\in\mathbb{R}^{3N}$ with unit length constraints $\|\mathbf m_i\|=1$.
We write the Hamiltonian as $E(\mathbf m)$ to avoid clashes with the Hessian
$H(\mathbf m)=\nabla^2 E(\mathbf m)$.
Gradients $\nabla E(\mathbf m)$ are obtained by reverse-mode automatic differentiation
and forces are $\mathbf F(\mathbf m)=-\nabla E(\mathbf m)$, understood as tangent-projected in spin systems.

\paragraph{HVP operator mode.}
We do not assemble $H$, instead we expose the linear operator
$\mathbf v \mapsto H(\mathbf m)\,\mathbf v$ through Hessian–vector products computed by the
autodiff engine. Autodiff provides two primitive Jacobian contractions for a vector-valued map
$\mathbf g(\mathbf m)$ with Jacobian $J(\mathbf m)=\partial \mathbf g/\partial \mathbf m$:
the \emph{Jacobian–vector product} (JVP),
\[
\mathrm{JVP}_{\mathbf g}(\mathbf m;\mathbf v)\;=\;J(\mathbf m)\,\mathbf v,
\]
and the \emph{vector–Jacobian product} (VJP),
\[
\mathrm{VJP}_{\mathbf g}(\mathbf m;\mathbf v)\;=\;J(\mathbf m)^{\!\top}\,\mathbf v .
\]
For the scalar energy $E(\mathbf m)$ with gradient $\nabla E(\mathbf m)$ (obtained via a VJP), the Hessian–vector product
is the directional derivative of the gradient,
\[
H(\mathbf m)\,\mathbf v
\;=\;
\left.\frac{d}{d\varepsilon}\,\nabla E(\mathbf m+\varepsilon\,\mathbf v)\right|_{\varepsilon=0}
\;=\;
\mathrm{JVP}_{\nabla E}(\mathbf m;\mathbf v),
\]
and can also be obtained by composing JVP and VJP (e.g., forward-over-reverse or reverse-over-forward),
while retaining $O(N)$ memory. We use this operator form in the eigensolvers.

For eigenmode estimation we call CuPy eigensolvers in operator mode and request the smallest
$k$ eigenpairs to form $V_{\min}$, without reimplementing Lanczos or LOBPCG.
This design is backend agnostic, the HVP callback can be swapped between PyTorch and JAX with no
change to the outer loop, and any solver that returns the $k$ smallest eigenpairs of a symmetric
operator can be used in place of Lanczos or LOBPCG 
\paragraph{Explicit Hessian for benchmarking.}
For small systems we construct the full manifold Hessian by applying PyTorch’s higher-order
autodiff in Jacobian mode to the tangent-projected gradient $\nabla E(\mathbf m)$, then take its
Hermitian part and solve the resulting dense Hermitian eigenproblem to obtain the smallest $k$
eigenpairs. These minimum modes serve as accuracy and scaling references. For moderate and large
systems, production calculations use the HVP-operator pipeline with Krylov eigensolvers instead of
assembling the Hessian.

\paragraph{Outer ascent optimizers on $\mathbf F_{\mathrm{mmf}}$.}
The outer climb operates directly on the minimum-mode force $\mathbf F_{\mathrm{mmf}}(\mathbf m)$.
We employ a GPU implementation of LBFGS (LBGFG and LBFGS-B) in our CuPy-Pytorch stack, and a nonlinear conjugate-gradient (CG) variant for comparison.
The optimizer is a pluggable component: any first-order or quasi-Newton routine can replace it (e.g., Adam, trust-region or truncated-Newton) as the framework only requires access to the force and the minimum-mode directions.

\subsection{Computational setup}

Multi-objective optimization used NSGA-II as implemented in pyMOO.\cite{Deb2002NSGAII,Blank2020pymoo}
Atomistic spin dynamics were performed with UppASD.\cite{Skubic2008JPCM}
Ni(111) heptamer structures were set up and relaxed with the Atomic Simulation Environment (ASE) using Cartesian degrees of freedom.\cite{Larsen2017JPCM}

\section{Code availability}
The reported results were obtained using standard open-source libraries. The optimization scheme described in the Methods and Supplementary Information was implemented with automatic differentiation operators in PyTorch, the NSGA-II algorithm as provided in the pyMOO package \cite{Deb2002NSGAII,Blank2020pymoo}, atomistic spin dynamics simulations with UppASD \cite{Skubic2008JPCM}, and structural relaxations with the Atomic Simulation Environment (ASE) \cite{Larsen2017JPCM}. The full algorithmic workflow is explicitly provided in the manuscript, which allows independent re-implementation using these widely available packages.

\section{Data availability}
All data supporting the findings of this study are available within the paper and its Supplementary Information.

\section{acknowledgments}
The authors thank Filipp N. Rybakov (Uppsala University), Pavel Bessarab (Linnaeus University) and Mathias Augustin (Uppsala University) for many fruitful discussions. We also thank Johan Hellsvik (KTH, PDC Center for High Performance Computing) for his support with GPU resources. The authors used AI-assisted tools to improve the language of the manuscript.

Financial support from the
Swedish Research Council (Vetenskapsr{\aa}det, VR) Grant No. 2016-05980, Grant No. 2019-05304, and Grant No. 2024-04986, and the Knut and Alice Wallenberg foundation Grant No. 2018.0060, Grant No. 2021.0246, and Grant No. 2022.0108 is acknowledged.  
The Wallenberg Initiative Materials Science for Sustainability (WISE) funded by the Knut and Alice Wallenberg Foundation is also acknowledged.
The computations/data handling were enabled by resources provided by the National Academic Infrastructure for Supercomputing in Sweden (NAISS), partially funded by the Swedish Research Council through grant agreement no. 2022-06725.

\section{Author Contributions}

Q.C. conceived the idea, carried out the research, A.D. supervised the project. Both authors contributed to the interpretation of the results and to the writing and revision of the manuscript.

\section{Competing Interests}
All authors declare that they have no conflicts of interest.
\bibliography{SA}

\section{Supplementary Information}
\subsection{Supplementary Note 1}

\begin{algorithm}
\caption{MOTO: Multi-objective three-layer minimum-mode-following optimization}
\label{alg:moto}
\begin{algorithmic}[1]
\Require reference minimum $x_0$, step rules, thresholds $(\tau_F,\tau_\lambda)$,
population size $P$, generations $G$
\State Initialize archive $\mathcal{A}\gets\varnothing$, population $\mathcal{S}\gets\{x_0+P_\Omega\,\delta_i\}_{i=1}^P$
\For{$g=1,\dots,G$}
  \For{$x\in\mathcal{S}$}
    \State $(\lambda_1,\lambda_2)\leftarrow \textsc{HVPOptmization}(x)$
    \State Evaluate objectives: diversity vs.\ $\mathcal{S}\cup\mathcal{A}$; physical constraint; spectral gap $\Delta\lambda=\lambda_2-\lambda_1$
  \EndFor
  \State $\mathcal{P}\leftarrow \textsc{NSGAIISortSelect}(\mathcal{S})$ \Comment{Pareto front}
  \For{$x\in\mathcal{P}$} \Comment{Bilayer MMF kernel}
    \Repeat
      \State $\mathbf{v}_{\min}\leftarrow \textsc{MinMode}(x)$ \Comment{HVP–Lanczos/LOBPCG in the masked tangent subspace}
      \State $\mathbf{F}\leftarrow -\nabla E(x)$
      \State $\mathbf{F}_{\mathrm{mmf}}\leftarrow \mathbf{F}-2(\mathbf{F}^{\!\top}\mathbf{v}_{\min})\mathbf{v}_{\min}$
      \State $x\leftarrow \mathcal{R}_{x}\!\bigl(x+\alpha\,\mathbf{F}_{\mathrm{mmf}}\bigr)$ \Comment{Lbfgs/CG, $\mathcal{R}$ is identity for atoms, retraction for spins}
    \Until{$\|\mathbf{F}_{\mathrm{mmf}}\|\le\tau_F$ \textbf{and} $\lambda_1<0<\lambda_2$}
    \State $x^\ddagger\gets x$ \Comment{index-1 saddle}
    \For{$s\in\{+1,-1\}$} \Comment{two-sided descent}
       \State $x_{\mathrm{seed}}\gets x^\ddagger+s\,\eta\,\mathbf{v}_{\min}(x^\ddagger)$
       \State $x_{\min}\leftarrow \textsc{RelaxDownhill}(x_{\mathrm{seed}})$
       \State \textsc{RegisterEdge}$(x_0,x^\ddagger,x_{\min},\mathcal{A})$
    \EndFor
  \EndFor
  \State $\mathcal{S}\leftarrow \textsc{EvolvePopulation}(\mathcal{S},\mathcal{A})$ \Comment{variation + archive feedback}
  \If{\textsc{CoverageStalled}$(\mathcal{A})$} \textbf{break} \EndIf
\EndFor
\State \Return local transition-state network encoded by $\mathcal{A}$
\end{algorithmic}
\end{algorithm}

\paragraph{Notation.}
Let $E(x)$ be the energy with configuration $x\in\mathbb{R}^N$, the force is
$\mathbf{F}(x)=-\nabla E(x)$ and the Hessian is $H(x)=\nabla^2 E(x)$, accessed through
Hessian–vector products $v\mapsto H(x)v$. The unit minimum-mode direction is
$\mathbf{v}_{\min}(x)$. For near-degenerate cases, a $k$-dimensional minimum-eigenspace
$V_{\min}(x)\in\mathbb{R}^{N\times k}$ with projector $P_{\min}(x)=V_{\min}(x)V_{\min}(x)^{\!\top}$
may be used.

\paragraph{Three layers.}
(i) \textbf{Multi-objective initial-configuration explorer.}
We evolve a population of masked perturbations of a reference minimum using NSGA-II–style selection
to optimize three objectives: diversity, feasibility, and minimum-mode separability
(the spectral gap $\Delta\lambda=\lambda_2-\lambda_1$ estimated via an HVP-based optmizer).
The output is a Pareto front of initial configurations.

(ii) \textbf{Bilayer MMF kernel.}
Inner layer: estimate $\mathbf{v}_{\min}(x)$ in the masked tangent subspace by minimizing the
Rayleigh quotient with HVP-based Lanczos/LOBPCG.
Outer layer: climb with the minimum-mode force used in the main text,
\[
\mathbf{F}_{\mathrm{mmf}}(x)\;=\;
\mathbf{F}(x)\;-\;2\bigl[\mathbf{F}(x)^{\!\top}\mathbf{v}_{\min}(x)\bigr]\mathbf{v}_{\min}(x),
\]
for spins, a retraction returns to the manifold.
Convergence at a saddle is declared when $\|\mathbf{F}_{\mathrm{mm}}(x)\|\le \tau_F$ and the local
curvature signature satisfies $\lambda_1(x)<0<\lambda_2(x)$ (estimated by the inner layer).

(iii) \textbf{Two-sided minimum-mode descent.}
From the saddle $x^\ddagger$, generate $x^\ddagger\pm\eta\,\mathbf{v}_{\min}(x^\ddagger)$ and relax
downhill to local minima using a deterministic optimizer (e.g., L-BFGS/CG) or dynamic simulator (e.g., UppASD for atomistic spin system). Outcomes are deduplicated
and added to the local network as certified edges.

\paragraph{Coverage loop and stopping.}
Repeat “explore $\rightarrow$ Bilayer MMF $\rightarrow$ certify” for the current Pareto set, successful configurations are archived and fed back to the diversity pool. Stop when no new saddles or minima appear for $K$ rounds, or when coverage metrics (unique saddles per generation, spectral-gap improvement) fall below a threshold. 

\subsection{Supplementary Note 2}
\paragraph{Supporting movies}
Supporting Movies~1–12 correspond to transition states TS1–TS12 in Fig.~3. Each movie visualizes
the bidirectional minimum-mode descent from the indexed saddle: the trajectory is initialized by
displacing the saddle configuration by $\pm\,\eta\,\mathbf{v}_{\min}$ and then relaxed with
atomistic spin dynamics at temperature $T=10^{-6}$ (internal units) and Gilbert damping
$\alpha=1$ for 50\,ps. The two branches (forward/backward) are shown to their respective terminal
minima, which are used to certify connectivity in the local transition-state network.

\subsection{Supplementary Note 3}
\paragraph{Complexity and scalability}

The Hessian-vector product offers several practical and computational advantages in the context of high-dimensional spin systems. First, it eliminates the need to explicitly construct or store the full Hessian matrix, which would otherwise require $\mathcal{O}(N^2)$ memory and $\mathcal{O}(N^3)$ computation for full eigendecomposition. Instead, the use of Hessian-vector products reduces both the memory and time complexity of the eigenvector estimation step to $\mathcal{O}(N)$ per HVP.

Each evaluation of the Rayleigh quotient gradient involves one HVP, which, in reverse-mode automatic differentiation, can be computed with nearly the same cost as evaluating a single gradient of the energy function. This is made possible by modern autograd systems like PyTorch, which construct and traverse the computational graph efficiently using vectorized chain rule applications. Thus, a single inner iteration of the eigenvector estimation loop has the same asymptotic cost as one backward pass.

Furthermore, the entire inner optimization procedure—typically 50 to 200 Adam steps—is independent of the system size in practice, since convergence is driven by local curvature rather than global dimensionality. For systems with tens or even hundreds of thousands of degrees of freedom, the total cost of estimating the minimum mode using this method remains orders of magnitude lower than computing the full Hessian and performing an exact diagonalization.

Importantly, this approach also enables on-the-fly optimization and GPU acceleration with minimal implementation overhead, as it reuses standard autograd primitives and optimizer infrastructure. Overall, the method is scalable, efficient, and well-suited to modern machine-learning-compatible spin simulation frameworks.

\subsection{Supplementary Note 4}
\paragraph{Interaction parameters}
Unless otherwise noted, the interaction parameters used in this work are
\textbf{reproduced from} Ref.~\cite{miranda2022band,xu2023metaheuristic,xu2025design,xu2022genetic}. Specifically,
Tables~\ref{tab:si_exchange_8col} and \ref{tab:si_dmi_8col} list the pairwise
Heisenberg exchange $J_{ij}$ and the Dzyaloshinskii–Moriya vectors
$\mathbf D_{ij}=(D_x,D_y,D_z)$, respectively, sorted by the intersite separation
$|\mathbf r_{ij}|$. No re-fitting was performed and values are rounded to six decimals for display.

    \begin{longtable*}{rrrrrrrr}
    \caption{Exchange couplings $J_{ij}$ sorted by $|\mathbf r_{ij}|$ (four pairs per row).}\label{tab:si_exchange_8col}\\
    \toprule
    \multicolumn{2}{c}{$J_{ij},\,|\mathbf r_{ij}|$} & \multicolumn{2}{c}{$J_{ij},\,|\mathbf r_{ij}|$} & \multicolumn{2}{c}{$J_{ij},\,|\mathbf r_{ij}|$} & \multicolumn{2}{c}{$J_{ij},\,|\mathbf r_{ij}|$} \\ \midrule
    $J_{ij}$ (mRy) & $|\mathbf r_{ij}|$ & $J_{ij}$ (mRy) & $|\mathbf r_{ij}|$ & $J_{ij}$ (mRy) & $|\mathbf r_{ij}|$ & $J_{ij}$ (mRy) & $|\mathbf r_{ij}|$ \\ \midrule
    \endfirsthead
    \toprule
    \multicolumn{2}{c}{$J_{ij},\,|\mathbf r_{ij}|$} & \multicolumn{2}{c}{$J_{ij},\,|\mathbf r_{ij}|$} & \multicolumn{2}{c}{$J_{ij},\,|\mathbf r_{ij}|$} & \multicolumn{2}{c}{$J_{ij},\,|\mathbf r_{ij}|$} \\ \midrule
    $J_{ij}$ (mRy) & $|\mathbf r_{ij}|$ & $J_{ij}$ (mRy) & $|\mathbf r_{ij}|$ & $J_{ij}$ (mRy) & $|\mathbf r_{ij}|$ & $J_{ij}$ (mRy) & $|\mathbf r_{ij}|$ \\ \midrule
    \endhead
    \midrule
    \endfoot
    \bottomrule
    \endlastfoot
    0.832467 & 0.707107 & 0.832467 & 0.707107 & 0.832467 & 0.707107 & 0.832467 & 0.707107 \\
0.832467 & 0.707107 & 0.832467 & 0.707107 & -0.007340 & 1.224745 & -0.007340 & 1.224745 \\
-0.007340 & 1.224745 & -0.007340 & 1.224745 & -0.007340 & 1.224745 & -0.007340 & 1.224745 \\
-0.157329 & 1.414214 & -0.157329 & 1.414214 & -0.157329 & 1.414214 & -0.157329 & 1.414214 \\
-0.157329 & 1.414214 & -0.157329 & 1.414214 & -0.020453 & 1.870829 & -0.020453 & 1.870829 \\
-0.020453 & 1.870829 & -0.020453 & 1.870829 & -0.020453 & 1.870829 & -0.020453 & 1.870829 \\
-0.020453 & 1.870829 & -0.020453 & 1.870829 & -0.020453 & 1.870829 & -0.020453 & 1.870829 \\
-0.020453 & 1.870829 & -0.020453 & 1.870829 & -0.015478 & 2.121320 & -0.015478 & 2.121320 \\
-0.015478 & 2.121320 & -0.015478 & 2.121320 & -0.015478 & 2.121320 & -0.015478 & 2.121320 \\
0.012835 & 2.449490 & 0.012835 & 2.449490 & 0.012835 & 2.449490 & 0.012835 & 2.449490 \\
0.012835 & 2.449490 & 0.012835 & 2.449490 & -0.002734 & 2.549510 & -0.002734 & 2.549510 \\
-0.002734 & 2.549510 & -0.002734 & 2.549510 & -0.002734 & 2.549510 & -0.002734 & 2.549510 \\
-0.002734 & 2.549510 & -0.002734 & 2.549510 & -0.002734 & 2.549510 & -0.002734 & 2.549510 \\
-0.002734 & 2.549510 & -0.002734 & 2.549510 & 0.006973 & 2.828427 & 0.006973 & 2.828427 \\
0.006973 & 2.828427 & 0.006973 & 2.828427 & 0.006973 & 2.828427 & 0.006973 & 2.828427 \\
-0.000746 & 3.082207 & -0.000746 & 3.082207 & -0.000746 & 3.082207 & -0.000746 & 3.082207 \\
-0.000746 & 3.082207 & -0.000746 & 3.082207 & -0.000746 & 3.082207 & -0.000746 & 3.082207 \\
-0.000746 & 3.082207 & -0.000746 & 3.082207 & -0.000746 & 3.082207 & -0.000746 & 3.082207 \\
0.002388 & 3.240370 & 0.002388 & 3.240370 & 0.002388 & 3.240370 & 0.002388 & 3.240370 \\
0.002388 & 3.240370 & 0.002388 & 3.240370 & 0.002388 & 3.240370 & 0.002388 & 3.240370 \\
0.002388 & 3.240370 & 0.002388 & 3.240370 & 0.002388 & 3.240370 & 0.002388 & 3.240370 \\
0.000760 & 3.535534 & 0.000760 & 3.535534 & 0.000760 & 3.535534 & 0.000760 & 3.535534 \\
0.000760 & 3.535534 & 0.000760 & 3.535534 & 0.000677 & 3.674235 & 0.000677 & 3.674235 \\
0.000677 & 3.674235 & 0.000677 & 3.674235 & 0.000677 & 3.674235 & 0.000677 & 3.674235 \\
-0.001129 & 3.741657 & -0.001129 & 3.741657 & -0.001129 & 3.741657 & -0.001129 & 3.741657 \\
-0.001129 & 3.741657 & -0.001129 & 3.741657 & -0.001129 & 3.741657 & -0.001129 & 3.741657 \\
-0.001129 & 3.741657 & -0.001129 & 3.741657 & -0.001129 & 3.741657 & -0.001129 & 3.741657 \\
0.000572 & 3.937004 & 0.000572 & 3.937004 & 0.000572 & 3.937004 & 0.000572 & 3.937004 \\
0.000572 & 3.937004 & 0.000572 & 3.937004 & 0.000572 & 3.937004 & 0.000572 & 3.937004 \\
0.000572 & 3.937004 & 0.000572 & 3.937004 & 0.000572 & 3.937004 & 0.000572 & 3.937004 \\
-0.000313 & 4.242641 & -0.000313 & 4.242641 & -0.000313 & 4.242641 & -0.000313 & 4.242641 \\
-0.000313 & 4.242641 & -0.000313 & 4.242641 & -0.000037 & 4.301163 & -0.000037 & 4.301163 \\
-0.000037 & 4.301163 & -0.000037 & 4.301163 & -0.000037 & 4.301163 & -0.000037 & 4.301163 \\
-0.000037 & 4.301163 & -0.000037 & 4.301163 & -0.000037 & 4.301163 & -0.000037 & 4.301163 \\
-0.000037 & 4.301163 & -0.000037 & 4.301163 & -0.000040 & 4.415880 & -0.000040 & 4.415880 \\
-0.000040 & 4.415880 & -0.000040 & 4.415880 & -0.000040 & 4.415880 & -0.000040 & 4.415880 \\
-0.000040 & 4.415880 & -0.000040 & 4.415880 & -0.000040 & 4.415880 & -0.000040 & 4.415880 \\
-0.000040 & 4.415880 & -0.000040 & 4.415880 & 0.000027 & 4.636809 & 0.000027 & 4.636809 \\
0.000027 & 4.636809 & 0.000027 & 4.636809 & 0.000027 & 4.636809 & 0.000027 & 4.636809 \\
0.000027 & 4.636809 & 0.000027 & 4.636809 & 0.000027 & 4.636809 & 0.000027 & 4.636809 \\
0.000027 & 4.636809 & 0.000027 & 4.636809 & 0.000043 & 4.898979 & 0.000043 & 4.898979 \\
0.000043 & 4.898979 & 0.000043 & 4.898979 & 0.000043 & 4.898979 & 0.000043 & 4.898979 \\
0.000008 & 4.949747 & -0.000084 & 4.949747 & 0.000008 & 4.949747 & 0.000008 & 4.949747 \\
-0.000084 & 4.949747 & 0.000008 & 4.949747 & 0.000008 & 4.949747 & -0.000084 & 4.949747 \\
0.000008 & 4.949747 & 0.000008 & 4.949747 & -0.000084 & 4.949747 & 0.000008 & 4.949747 \\
0.000008 & 4.949747 & -0.000084 & 4.949747 & 0.000008 & 4.949747 & 0.000008 & 4.949747 \\
-0.000084 & 4.949747 & 0.000008 & 4.949747 & 0.000042 & 5.099020 & 0.000042 & 5.099020 \\
0.000042 & 5.099020 & 0.000042 & 5.099020 & 0.000042 & 5.099020 & 0.000042 & 5.099020 \\
0.000042 & 5.099020 & 0.000042 & 5.099020 & 0.000042 & 5.099020 & 0.000042 & 5.099020 \\
0.000042 & 5.099020 & 0.000042 & 5.099020 & 0.000086 & 5.338539 & 0.000086 & 5.338539 \\
0.000086 & 5.338539 & 0.000086 & 5.338539 & 0.000086 & 5.338539 & 0.000086 & 5.338539 \\
0.000086 & 5.338539 & 0.000086 & 5.338539 & 0.000086 & 5.338539 & 0.000086 & 5.338539 \\
0.000086 & 5.338539 & 0.000086 & 5.338539 & 0.000037 & 5.522681 & 0.000037 & 5.522681 \\
0.000037 & 5.522681 & 0.000037 & 5.522681 & 0.000037 & 5.522681 & 0.000037 & 5.522681 \\
0.000037 & 5.522681 & 0.000037 & 5.522681 & 0.000037 & 5.522681 & 0.000037 & 5.522681 \\
0.000037 & 5.522681 & 0.000037 & 5.522681 & 0.000005 & 5.612486 & 0.000005 & 5.612486 \\
0.000005 & 5.612486 & 0.000005 & 5.612486 & 0.000005 & 5.612486 & 0.000005 & 5.612486 \\
0.000005 & 5.612486 & 0.000005 & 5.612486 & 0.000005 & 5.612486 & 0.000005 & 5.612486 \\
0.000005 & 5.612486 & 0.000005 & 5.612486 & 0.000028 & 5.656854 & 0.000028 & 5.656854 \\
0.000028 & 5.656854 & 0.000028 & 5.656854 & 0.000028 & 5.656854 & 0.000028 & 5.656854 \\
0.000018 & 5.787918 & 0.000018 & 5.787918 & 0.000018 & 5.787918 & 0.000018 & 5.787918 \\
0.000018 & 5.787918 & 0.000018 & 5.787918 & 0.000018 & 5.787918 & 0.000018 & 5.787918 \\
0.000018 & 5.787918 & 0.000018 & 5.787918 & 0.000018 & 5.787918 & 0.000018 & 5.787918 \\
0.000005 & 6.041523 & 0.000005 & 6.041523 & 0.000005 & 6.041523 & 0.000005 & 6.041523 \\
0.000005 & 6.041523 & 0.000005 & 6.041523 & 0.000005 & 6.041523 & 0.000005 & 6.041523 \\
0.000005 & 6.041523 & 0.000005 & 6.041523 & 0.000005 & 6.041523 & 0.000005 & 6.041523 \\
-0.000001 & 6.123724 & -0.000001 & 6.123724 & -0.000000 & 6.123724 & -0.000001 & 6.123724 \\
0.000000 & 6.123724 & -0.000000 & 6.123724 & -0.000023 & 6.164414 & -0.000023 & 6.164414 \\
-0.000023 & 6.164414 & -0.000023 & 6.164414 & -0.000023 & 6.164414 & -0.000023 & 6.164414 \\
-0.000023 & 6.164414 & -0.000023 & 6.164414 & -0.000023 & 6.164414 & -0.000023 & 6.164414 \\
-0.000023 & 6.164414 & -0.000023 & 6.164414 & 0.000009 & 6.284903 & 0.000009 & 6.284903 \\
0.000009 & 6.284903 & 0.000009 & 6.284903 & 0.000009 & 6.284903 & 0.000010 & 6.284903 \\
0.000009 & 6.284903 & 0.000009 & 6.284903 & 0.000010 & 6.284903 & 0.000010 & 6.284903 \\
0.000009 & 6.284903 & 0.000010 & 6.284903 & 0.000024 & 6.363961 & 0.000023 & 6.363961 \\
0.000024 & 6.363961 & 0.000023 & 6.363961 & 0.000023 & 6.363961 & 0.000023 & 6.363961 \\
-0.000021 & 6.480741 & -0.000021 & 6.480741 & -0.000021 & 6.480741 & -0.000021 & 6.480741 \\
-0.000021 & 6.480741 & -0.000021 & 6.480741 & -0.000022 & 6.480741 & -0.000021 & 6.480741 \\
-0.000021 & 6.480741 & -0.000021 & 6.480741 & -0.000021 & 6.480741 & -0.000022 & 6.480741 \\
-0.000003 & 6.745369 & -0.000011 & 6.745369 & -0.000011 & 6.745369 & -0.000003 & 6.745369 \\
-0.000004 & 6.745369 & -0.000011 & 6.745369 & -0.000011 & 6.745369 & -0.000003 & 6.745369 \\
-0.000003 & 6.745369 & -0.000009 & 6.745369 & -0.000009 & 6.745369 & -0.000003 & 6.745369 \\
-0.000003 & 6.745369 & -0.000011 & 6.745369 & -0.000011 & 6.745369 & -0.000004 & 6.745369 \\
-0.000004 & 6.745369 & -0.000010 & 6.745369 & -0.000010 & 6.745369 & -0.000003 & 6.745369 \\
-0.000004 & 6.745369 & -0.000009 & 6.745369 & -0.000009 & 6.745369 & -0.000003 & 6.745369 \\
-0.000013 & 6.819091 & -0.000012 & 6.819091 & -0.000013 & 6.819091 & -0.000013 & 6.819091 \\
-0.000011 & 6.819091 & -0.000010 & 6.819091 & -0.000013 & 6.819091 & -0.000012 & 6.819091 \\
-0.000011 & 6.819091 & -0.000011 & 6.819091 & -0.000011 & 6.819091 & -0.000010 & 6.819091 \\
-0.000003 & 6.964194 & -0.000003 & 6.964194 & -0.000003 & 6.964194 & -0.000003 & 6.964194 \\
-0.000002 & 6.964194 & -0.000001 & 6.964194 & -0.000003 & 6.964194 & -0.000003 & 6.964194 \\
-0.000002 & 6.964194 & -0.000002 & 6.964194 & -0.000002 & 6.964194 & -0.000001 & 6.964194 \\
    \end{longtable*}

\begin{longtable*}{rrrrrrrr}
    \caption{Dzyaloshinskii-Moriya vectors sorted by $|\mathbf r_{ij}|$ (two bonds per row).}\label{tab:si_dmi_8col}\\
    \toprule
    \multicolumn{4}{c}{$\mathbf D_{ij}=(D_x,D_y,D_z),\,|\mathbf r_{ij}|$} & \multicolumn{4}{c}{$\mathbf D_{ij}=(D_x,D_y,D_z),\,|\mathbf r_{ij}|$} \\ \midrule
    $D_x$ (mRy) & $D_y$ (mRy) & $D_z$ (mRy) & $|\mathbf r_{ij}|$ & $D_x$ (mRy) & $D_y$ (mRy) & $D_z$ (mRy) & $|\mathbf r_{ij}|$ \\ \midrule
    \endfirsthead
    \toprule
    \multicolumn{4}{c}{$\mathbf D_{ij}=(D_x,D_y,D_z),\,|\mathbf r_{ij}|$} & \multicolumn{4}{c}{$\mathbf D_{ij}=(D_x,D_y,D_z),\,|\mathbf r_{ij}|$} \\ \midrule
    $D_x$ (m) & $D_y$ (mRy) & $D_z$ (mRy) & $|\mathbf r_{ij}|$ & $D_x$ (mRy) & $D_y$ (mRy) & $D_z$ (mRy) & $|\mathbf r_{ij}|$ \\ \midrule
    \endhead
    \midrule
    \endfoot
    \bottomrule
    \endlastfoot
    0.010523 & -0.039268 & -0.020318 & 0.707107 & -0.028745 & -0.028747 & 0.020318 & 0.707107 \\
0.039268 & -0.010521 & 0.020318 & 0.707107 & -0.039268 & 0.010521 & -0.020318 & 0.707107 \\
0.028745 & 0.028747 & -0.020318 & 0.707107 & -0.010523 & 0.039268 & 0.020318 & 0.707107 \\
0.003254 & 0.012145 & -0.000000 & 1.224745 & -0.008891 & 0.008891 & 0.000000 & 1.224745 \\
0.012145 & 0.003255 & 0.000000 & 1.224745 & -0.012145 & -0.003255 & -0.000000 & 1.224745 \\
0.008891 & -0.008891 & -0.000000 & 1.224745 & -0.003254 & -0.012145 & 0.000000 & 1.224745 \\
-0.001992 & 0.007435 & 0.000621 & 1.414214 & 0.005443 & 0.005443 & -0.000621 & 1.414214 \\
-0.007435 & 0.001992 & -0.000621 & 1.414214 & 0.007435 & -0.001992 & 0.000621 & 1.414214 \\
-0.005443 & -0.005443 & 0.000621 & 1.414214 & 0.001992 & -0.007435 & -0.000621 & 1.414214 \\
0.000960 & -0.004895 & -0.000614 & 1.870829 & -0.003279 & -0.003759 & 0.000614 & 1.870829 \\
0.001616 & -0.004719 & -0.000614 & 1.870829 & 0.004719 & -0.001616 & 0.000614 & 1.870829 \\
-0.003759 & -0.003279 & 0.000614 & 1.870829 & -0.004895 & 0.000960 & -0.000614 & 1.870829 \\
0.004895 & -0.000960 & 0.000614 & 1.870829 & 0.003759 & 0.003279 & -0.000614 & 1.870829 \\
-0.004719 & 0.001616 & -0.000614 & 1.870829 & -0.001616 & 0.004719 & 0.000614 & 1.870829 \\
0.003279 & 0.003759 & -0.000614 & 1.870829 & -0.000960 & 0.004895 & 0.000614 & 1.870829 \\
0.001643 & -0.006131 & 0.001329 & 2.121320 & -0.004488 & -0.004488 & -0.001329 & 2.121320 \\
0.006131 & -0.001643 & -0.001329 & 2.121320 & -0.006131 & 0.001643 & 0.001329 & 2.121320 \\
0.004488 & 0.004488 & 0.001329 & 2.121320 & -0.001643 & 0.006131 & -0.001329 & 2.121320 \\
0.000187 & 0.000698 & 0.000000 & 2.449490 & -0.000511 & 0.000511 & -0.000000 & 2.449490 \\
0.000698 & 0.000187 & -0.000000 & 2.449490 & -0.000698 & -0.000187 & 0.000000 & 2.449490 \\
0.000511 & -0.000511 & 0.000000 & 2.449490 & -0.000187 & -0.000698 & -0.000000 & 2.449490 \\
-0.000773 & 0.000693 & 0.000575 & 2.549510 & 0.000323 & 0.000987 & 0.000575 & 2.549510 \\
0.001016 & 0.000214 & -0.000575 & 2.549510 & 0.000214 & 0.001016 & -0.000575 & 2.549510 \\
-0.000987 & -0.000323 & -0.000575 & 2.549510 & -0.000693 & 0.000773 & -0.000575 & 2.549510 \\
0.000693 & -0.000773 & 0.000575 & 2.549510 & 0.000987 & 0.000323 & 0.000575 & 2.549510 \\
-0.000214 & -0.001016 & 0.000575 & 2.549510 & -0.001016 & -0.000214 & 0.000575 & 2.549510 \\
-0.000323 & -0.000987 & -0.000575 & 2.549510 & 0.000773 & -0.000693 & -0.000575 & 2.549510 \\
-0.000195 & 0.000729 & -0.000821 & 2.828427 & 0.000534 & 0.000534 & 0.000821 & 2.828427 \\
-0.000729 & 0.000195 & 0.000821 & 2.828427 & 0.000729 & -0.000195 & -0.000821 & 2.828427 \\
-0.000534 & -0.000534 & -0.000821 & 2.828427 & 0.000195 & -0.000729 & 0.000821 & 2.828427 \\
-0.000079 & -0.000657 & -0.000206 & 3.082207 & -0.000260 & -0.000609 & 0.000206 & 3.082207 \\
0.000397 & -0.000530 & -0.000206 & 3.082207 & 0.000530 & -0.000397 & 0.000206 & 3.082207 \\
-0.000609 & -0.000260 & 0.000206 & 3.082207 & -0.000657 & -0.000079 & -0.000206 & 3.082207 \\
0.000657 & 0.000079 & 0.000206 & 3.082207 & 0.000609 & 0.000260 & -0.000206 & 3.082207 \\
-0.000530 & 0.000397 & -0.000206 & 3.082207 & -0.000397 & 0.000530 & 0.000206 & 3.082207 \\
0.000260 & 0.000609 & -0.000206 & 3.082207 & 0.000079 & 0.000657 & 0.000206 & 3.082207 \\
-0.000059 & -0.000197 & -0.000160 & 3.240370 & 0.000150 & -0.000141 & -0.000160 & 3.240370 \\
-0.000047 & -0.000200 & 0.000160 & 3.240370 & -0.000200 & -0.000047 & 0.000160 & 3.240370 \\
0.000141 & -0.000150 & 0.000160 & 3.240370 & 0.000197 & 0.000059 & 0.000160 & 3.240370 \\
-0.000197 & -0.000059 & -0.000160 & 3.240370 & -0.000141 & 0.000150 & -0.000160 & 3.240370 \\
0.000200 & 0.000047 & -0.000160 & 3.240370 & 0.000047 & 0.000200 & -0.000160 & 3.240370 \\
-0.000150 & 0.000141 & 0.000160 & 3.240370 & 0.000059 & 0.000197 & 0.000160 & 3.240370 \\
0.000048 & -0.000179 & 0.000047 & 3.535534 & -0.000131 & -0.000131 & -0.000047 & 3.535534 \\
0.000179 & -0.000048 & -0.000047 & 3.535534 & -0.000179 & 0.000048 & 0.000047 & 3.535534 \\
0.000131 & 0.000131 & 0.000047 & 3.535534 & -0.000048 & 0.000179 & -0.000047 & 3.535534 \\
0.000012 & 0.000043 & -0.000000 & 3.674235 & -0.000031 & 0.000031 & 0.000000 & 3.674235 \\
0.000043 & 0.000012 & 0.000000 & 3.674235 & -0.000043 & -0.000012 & -0.000000 & 3.674235 \\
0.000031 & -0.000031 & -0.000000 & 3.674235 & -0.000012 & -0.000043 & 0.000000 & 3.674235 \\
-0.000034 & -0.000309 & 0.000123 & 3.741657 & -0.000125 & -0.000285 & -0.000123 & 3.741657 \\
0.000184 & -0.000251 & 0.000123 & 3.741657 & 0.000251 & -0.000184 & -0.000123 & 3.741657 \\
-0.000285 & -0.000125 & -0.000123 & 3.741657 & -0.000309 & -0.000034 & 0.000123 & 3.741657 \\
0.000309 & 0.000034 & -0.000123 & 3.741657 & 0.000285 & 0.000125 & 0.000123 & 3.741657 \\
-0.000251 & 0.000184 & 0.000123 & 3.741657 & -0.000184 & 0.000251 & -0.000123 & 3.741657 \\
0.000125 & 0.000285 & 0.000123 & 3.741657 & 0.000034 & 0.000309 & -0.000123 & 3.741657 \\
-0.000082 & 0.000177 & -0.000036 & 3.937004 & -0.000018 & 0.000195 & -0.000036 & 3.937004 \\
0.000160 & 0.000113 & 0.000036 & 3.937004 & 0.000113 & 0.000160 & 0.000036 & 3.937004 \\
-0.000195 & 0.000018 & 0.000036 & 3.937004 & -0.000177 & 0.000082 & 0.000036 & 3.937004 \\
0.000177 & -0.000082 & -0.000036 & 3.937004 & 0.000195 & -0.000018 & -0.000036 & 3.937004 \\
-0.000113 & -0.000160 & -0.000036 & 3.937004 & -0.000160 & -0.000113 & -0.000036 & 3.937004 \\
0.000018 & -0.000195 & 0.000036 & 3.937004 & 0.000082 & -0.000177 & 0.000036 & 3.937004 \\
0.000092 & -0.000342 & 0.000074 & 4.242641 & -0.000251 & -0.000251 & -0.000074 & 4.242641 \\
0.000342 & -0.000092 & -0.000074 & 4.242641 & -0.000342 & 0.000092 & 0.000074 & 4.242641 \\
0.000251 & 0.000251 & 0.000074 & 4.242641 & -0.000092 & 0.000342 & -0.000074 & 4.242641 \\
0.000085 & 0.000074 & -0.000055 & 4.301163 & -0.000037 & 0.000107 & 0.000055 & 4.301163 \\
-0.000111 & 0.000022 & -0.000055 & 4.301163 & -0.000022 & 0.000111 & 0.000055 & 4.301163 \\
0.000107 & -0.000037 & 0.000055 & 4.301163 & 0.000074 & 0.000085 & -0.000055 & 4.301163 \\
-0.000074 & -0.000085 & 0.000055 & 4.301163 & -0.000107 & 0.000037 & -0.000055 & 4.301163 \\
0.000022 & -0.000111 & -0.000055 & 4.301163 & 0.000111 & -0.000022 & 0.000055 & 4.301163 \\
0.000037 & -0.000107 & -0.000055 & 4.301163 & -0.000085 & -0.000074 & 0.000055 & 4.301163 \\
0.000021 & -0.000026 & -0.000059 & 4.415880 & -0.000032 & -0.000012 & 0.000059 & 4.415880 \\
-0.000005 & -0.000034 & -0.000059 & 4.415880 & 0.000034 & 0.000005 & 0.000059 & 4.415880 \\
-0.000012 & -0.000032 & 0.000059 & 4.415880 & -0.000026 & 0.000021 & -0.000059 & 4.415880 \\
0.000026 & -0.000021 & 0.000059 & 4.415880 & 0.000012 & 0.000032 & -0.000059 & 4.415880 \\
-0.000034 & -0.000005 & -0.000059 & 4.415880 & 0.000005 & 0.000034 & 0.000059 & 4.415880 \\
0.000032 & 0.000012 & -0.000059 & 4.415880 & -0.000021 & 0.000026 & 0.000059 & 4.415880 \\
-0.000029 & 0.000145 & 0.000059 & 4.636809 & -0.000047 & 0.000140 & 0.000059 & 4.636809 \\
0.000098 & 0.000111 & -0.000059 & 4.636809 & 0.000111 & 0.000098 & -0.000059 & 4.636809 \\
-0.000140 & 0.000047 & -0.000059 & 4.636809 & -0.000145 & 0.000029 & -0.000059 & 4.636809 \\
0.000145 & -0.000029 & 0.000059 & 4.636809 & 0.000140 & -0.000047 & 0.000059 & 4.636809 \\
-0.000111 & -0.000098 & 0.000059 & 4.636809 & -0.000098 & -0.000111 & 0.000059 & 4.636809 \\
0.000047 & -0.000140 & -0.000059 & 4.636809 & 0.000029 & -0.000145 & -0.000059 & 4.636809 \\
0.000014 & 0.000052 & -0.000000 & 4.898979 & -0.000038 & 0.000038 & 0.000000 & 4.898979 \\
0.000052 & 0.000014 & 0.000000 & 4.898979 & -0.000052 & -0.000014 & -0.000000 & 4.898979 \\
0.000038 & -0.000038 & -0.000000 & 4.898979 & -0.000014 & -0.000052 & 0.000000 & 4.898979 \\
-0.000023 & 0.000014 & -0.000022 & 4.949747 & 0.000021 & -0.000080 & -0.000048 & 4.949747 \\
0.000027 & 0.000001 & 0.000022 & 4.949747 & 0.000012 & 0.000024 & -0.000022 & 4.949747 \\
-0.000058 & -0.000058 & 0.000048 & 4.949747 & -0.000024 & -0.000012 & 0.000022 & 4.949747 \\
0.000001 & 0.000027 & 0.000022 & 4.949747 & 0.000080 & -0.000021 & 0.000048 & 4.949747 \\
0.000014 & -0.000023 & -0.000022 & 4.949747 & -0.000014 & 0.000023 & 0.000022 & 4.949747 \\
-0.000080 & 0.000021 & -0.000048 & 4.949747 & -0.000001 & -0.000027 & -0.000022 & 4.949747 \\
0.000024 & 0.000012 & -0.000022 & 4.949747 & 0.000058 & 0.000058 & -0.000048 & 4.949747 \\
-0.000012 & -0.000024 & 0.000022 & 4.949747 & -0.000027 & -0.000001 & -0.000022 & 4.949747 \\
-0.000021 & 0.000080 & 0.000048 & 4.949747 & 0.000023 & -0.000014 & 0.000022 & 4.949747 \\
0.000015 & -0.000046 & -0.000004 & 5.099020 & 0.000010 & -0.000048 & -0.000004 & 5.099020 \\
-0.000036 & -0.000033 & 0.000004 & 5.099020 & -0.000033 & -0.000036 & 0.000004 & 5.099020 \\
0.000048 & -0.000010 & 0.000004 & 5.099020 & 0.000046 & -0.000015 & 0.000004 & 5.099020 \\
-0.000046 & 0.000015 & -0.000004 & 5.099020 & -0.000048 & 0.000010 & -0.000004 & 5.099020 \\
0.000033 & 0.000036 & -0.000004 & 5.099020 & 0.000036 & 0.000033 & -0.000004 & 5.099020 \\
-0.000010 & 0.000048 & 0.000004 & 5.099020 & -0.000015 & 0.000046 & 0.000004 & 5.099020 \\
0.000016 & 0.000019 & 0.000047 & 5.338539 & -0.000023 & 0.000009 & 0.000047 & 5.338539 \\
-0.000004 & 0.000025 & -0.000047 & 5.338539 & 0.000025 & -0.000004 & -0.000047 & 5.338539 \\
-0.000009 & 0.000023 & -0.000047 & 5.338539 & -0.000019 & -0.000016 & -0.000047 & 5.338539 \\
0.000019 & 0.000016 & 0.000047 & 5.338539 & 0.000009 & -0.000023 & 0.000047 & 5.338539 \\
-0.000025 & 0.000004 & 0.000047 & 5.338539 & 0.000004 & -0.000025 & 0.000047 & 5.338539 \\
0.000023 & -0.000009 & -0.000047 & 5.338539 & -0.000016 & -0.000019 & -0.000047 & 5.338539 \\
-0.000011 & -0.000027 & -0.000005 & 5.522681 & -0.000004 & -0.000029 & 0.000005 & 5.522681 \\
0.000023 & -0.000018 & -0.000005 & 5.522681 & 0.000018 & -0.000023 & 0.000005 & 5.522681 \\
-0.000029 & -0.000004 & 0.000005 & 5.522681 & -0.000027 & -0.000011 & -0.000005 & 5.522681 \\
0.000027 & 0.000011 & 0.000005 & 5.522681 & 0.000029 & 0.000004 & -0.000005 & 5.522681 \\
-0.000018 & 0.000023 & -0.000005 & 5.522681 & -0.000023 & 0.000018 & 0.000005 & 5.522681 \\
0.000004 & 0.000029 & -0.000005 & 5.522681 & 0.000011 & 0.000027 & 0.000005 & 5.522681 \\
0.000012 & -0.000002 & 0.000011 & 5.612486 & -0.000011 & 0.000004 & -0.000011 & 5.612486 \\
-0.000009 & -0.000008 & 0.000011 & 5.612486 & 0.000008 & 0.000009 & -0.000011 & 5.612486 \\
0.000004 & -0.000011 & -0.000011 & 5.612486 & -0.000002 & 0.000012 & 0.000011 & 5.612486 \\
0.000002 & -0.000012 & -0.000011 & 5.612486 & -0.000004 & 0.000011 & 0.000011 & 5.612486 \\
-0.000008 & -0.000009 & 0.000011 & 5.612486 & 0.000009 & 0.000008 & -0.000011 & 5.612486 \\
0.000011 & -0.000004 & 0.000011 & 5.612486 & -0.000012 & 0.000002 & -0.000011 & 5.612486 \\
-0.000004 & 0.000013 & 0.000011 & 5.656854 & 0.000010 & 0.000010 & -0.000011 & 5.656854 \\
-0.000013 & 0.000004 & -0.000011 & 5.656854 & 0.000013 & -0.000004 & 0.000011 & 5.656854 \\
-0.000010 & -0.000010 & 0.000011 & 5.656854 & 0.000004 & -0.000013 & -0.000011 & 5.656854 \\
-0.000011 & -0.000019 & -0.000006 & 5.787918 & 0.000019 & -0.000011 & -0.000006 & 5.787918 \\
-0.000000 & -0.000022 & 0.000006 & 5.787918 & -0.000022 & -0.000000 & 0.000006 & 5.787918 \\
0.000011 & -0.000019 & 0.000006 & 5.787918 & 0.000019 & 0.000011 & 0.000006 & 5.787918 \\
-0.000019 & -0.000011 & -0.000006 & 5.787918 & -0.000011 & 0.000019 & -0.000006 & 5.787918 \\
0.000022 & 0.000000 & -0.000006 & 5.787918 & 0.000000 & 0.000022 & -0.000006 & 5.787918 \\
-0.000019 & 0.000011 & 0.000006 & 5.787918 & 0.000011 & 0.000019 & 0.000006 & 5.787918 \\
0.000005 & 0.000008 & 0.000007 & 6.041523 & -0.000008 & 0.000005 & 0.000007 & 6.041523 \\
-0.000000 & 0.000010 & -0.000007 & 6.041523 & 0.000010 & -0.000000 & -0.000007 & 6.041523 \\
-0.000005 & 0.000008 & -0.000007 & 6.041523 & -0.000008 & -0.000005 & -0.000007 & 6.041523 \\
0.000008 & 0.000005 & 0.000007 & 6.041523 & 0.000005 & -0.000008 & 0.000007 & 6.041523 \\
-0.000010 & 0.000000 & 0.000007 & 6.041523 & 0.000000 & -0.000010 & 0.000007 & 6.041523 \\
0.000008 & -0.000005 & -0.000007 & 6.041523 & -0.000005 & -0.000008 & -0.000007 & 6.041523 \\
-0.000006 & -0.000024 & -0.000000 & 6.123724 & 0.000018 & -0.000018 & 0.000000 & 6.123724 \\
-0.000024 & -0.000006 & 0.000000 & 6.123724 & 0.000024 & 0.000006 & -0.000000 & 6.123724 \\
-0.000018 & 0.000018 & -0.000000 & 6.123724 & 0.000006 & 0.000024 & 0.000000 & 6.123724 \\
-0.000007 & -0.000014 & 0.000004 & 6.164414 & -0.000000 & -0.000016 & -0.000004 & 6.164414 \\
0.000013 & -0.000008 & 0.000004 & 6.164414 & 0.000008 & -0.000013 & -0.000004 & 6.164414 \\
-0.000016 & -0.000000 & -0.000004 & 6.164414 & -0.000014 & -0.000007 & 0.000004 & 6.164414 \\
0.000014 & 0.000007 & -0.000004 & 6.164414 & 0.000016 & 0.000000 & 0.000004 & 6.164414 \\
-0.000008 & 0.000013 & 0.000004 & 6.164414 & -0.000013 & 0.000008 & -0.000004 & 6.164414 \\
0.000001 & 0.000016 & 0.000004 & 6.164414 & 0.000007 & 0.000014 & -0.000004 & 6.164414 \\
-0.000013 & -0.000014 & -0.000001 & 6.284903 & 0.000004 & -0.000019 & 0.000001 & 6.284903 \\
0.000018 & -0.000005 & -0.000001 & 6.284903 & 0.000005 & -0.000018 & 0.000001 & 6.284903 \\
-0.000019 & 0.000005 & 0.000001 & 6.284903 & -0.000014 & -0.000013 & -0.000001 & 6.284903 \\
0.000014 & 0.000013 & 0.000001 & 6.284903 & 0.000019 & -0.000004 & -0.000001 & 6.284903 \\
-0.000006 & 0.000018 & -0.000001 & 6.284903 & -0.000018 & 0.000006 & 0.000001 & 6.284903 \\
-0.000005 & 0.000019 & -0.000001 & 6.284903 & 0.000013 & 0.000014 & 0.000001 & 6.284903 \\
-0.000003 & 0.000010 & -0.000006 & 6.363961 & 0.000007 & 0.000007 & 0.000005 & 6.363961 \\
-0.000010 & 0.000003 & 0.000006 & 6.363961 & 0.000010 & -0.000002 & -0.000005 & 6.363961 \\
-0.000007 & -0.000007 & -0.000005 & 6.363961 & 0.000002 & -0.000010 & 0.000005 & 6.363961 \\
0.000000 & -0.000004 & -0.000001 & 6.480741 & 0.000002 & -0.000003 & -0.000001 & 6.480741 \\
-0.000002 & -0.000003 & 0.000001 & 6.480741 & -0.000003 & -0.000002 & 0.000001 & 6.480741 \\
0.000003 & -0.000002 & 0.000001 & 6.480741 & 0.000004 & -0.000000 & 0.000001 & 6.480741 \\
-0.000004 & -0.000000 & -0.000001 & 6.480741 & -0.000003 & 0.000002 & -0.000001 & 6.480741 \\
0.000003 & 0.000002 & -0.000001 & 6.480741 & 0.000002 & 0.000003 & -0.000001 & 6.480741 \\
-0.000002 & 0.000003 & 0.000001 & 6.480741 & 0.000000 & 0.000004 & 0.000001 & 6.480741 \\
0.000001 & -0.000002 & -0.000002 & 6.745369 & 0.000006 & 0.000010 & 0.000000 & 6.745369 \\
0.000000 & 0.000012 & -0.000000 & 6.745369 & 0.000000 & -0.000003 & -0.000002 & 6.745369 \\
-0.000002 & -0.000002 & 0.000002 & 6.745369 & -0.000010 & 0.000006 & 0.000000 & 6.745369 \\
-0.000006 & 0.000010 & -0.000000 & 6.745369 & -0.000001 & -0.000002 & 0.000002 & 6.745369 \\
0.000003 & -0.000000 & 0.000002 & 6.745369 & 0.000013 & 0.000000 & -0.000000 & 6.745369 \\
0.000012 & 0.000006 & 0.000000 & 6.745369 & 0.000002 & -0.000001 & 0.000002 & 6.745369 \\
-0.000003 & 0.000001 & -0.000002 & 6.745369 & -0.000010 & -0.000006 & -0.000000 & 6.745369 \\
-0.000012 & -0.000000 & 0.000000 & 6.745369 & -0.000003 & 0.000001 & -0.000002 & 6.745369 \\
0.000002 & 0.000002 & -0.000002 & 6.745369 & 0.000007 & -0.000010 & 0.000000 & 6.745369 \\
0.000010 & -0.000007 & -0.000000 & 6.745369 & 0.000002 & 0.000001 & -0.000002 & 6.745369 \\
-0.000001 & 0.000003 & 0.000002 & 6.745369 & -0.000000 & -0.000013 & 0.000000 & 6.745369 \\
-0.000006 & -0.000012 & -0.000000 & 6.745369 & -0.000001 & 0.000003 & 0.000002 & 6.745369 \\
0.000005 & 0.000004 & 0.000006 & 6.819091 & -0.000002 & 0.000006 & -0.000006 & 6.819091 \\
-0.000006 & 0.000001 & 0.000006 & 6.819091 & -0.000001 & 0.000006 & -0.000006 & 6.819091 \\
0.000007 & -0.000002 & -0.000006 & 6.819091 & 0.000005 & 0.000005 & 0.000006 & 6.819091 \\
-0.000004 & -0.000005 & -0.000006 & 6.819091 & -0.000006 & 0.000002 & 0.000006 & 6.819091 \\
0.000002 & -0.000006 & 0.000006 & 6.819091 & 0.000006 & -0.000002 & -0.000006 & 6.819091 \\
0.000002 & -0.000007 & 0.000006 & 6.819091 & -0.000005 & -0.000005 & -0.000006 & 6.819091 \\
0.000004 & -0.000003 & 0.000002 & 6.964194 & -0.000005 & -0.000001 & -0.000002 & 6.964194 \\
-0.000002 & -0.000005 & 0.000003 & 6.964194 & 0.000005 & 0.000002 & -0.000003 & 6.964194 \\
-0.000001 & -0.000005 & -0.000002 & 6.964194 & -0.000003 & 0.000004 & 0.000002 & 6.964194 \\
0.000003 & -0.000004 & -0.000002 & 6.964194 & 0.000001 & 0.000005 & 0.000002 & 6.964194 \\
-0.000005 & -0.000002 & 0.000001 & 6.964194 & 0.000002 & 0.000005 & -0.000001 & 6.964194 \\
0.000005 & 0.000001 & 0.000002 & 6.964194 & -0.000004 & 0.000004 & -0.000002 & 6.964194 \\
    \end{longtable*}

\end{document}